\documentclass[preprint]{aastex}
\usepackage[utf8]{inputenc}
\usepackage{color}
\usepackage{booktabs}
\usepackage{natbib}
\usepackage{verbatim}
\usepackage{comment}

\newcommand{\thisevent}{OGLE-2015-BLG-0966}

\usepackage{xr}

\title{The Radial Velocity of OGLE-2015-BLG-0966S}
\author{Samson A. Johnson\altaffilmark{1},
Jennifer C. Yee\altaffilmark{2}
\altaffiltext{1}{Dept. of Astronomy, Ohio State University, 140 W
18th Ave, Columbus, OH 43210, USA, johnson.7080@osu.edu}
\altaffiltext{2}{Smithsonian Astrophysical Observatory, 60 Garden
Street, Cambridge, MA 02138, USA}
}

\usepackage{amsmath}
\usepackage{bm}

\usepackage{hyperref}
\hypersetup{
    colorlinks=true,
    linkcolor=blue,
    filecolor=magenta,      
    urlcolor=cyan,
}

\begin{document}

\begin{abstract}
    The distance to the planetary system \thisevent L and the separation between the planet and its host star are ambiguous due to an ambiguity in the distance to the source star \citep{Street16}. We attempt to break this degeneracy by measuring the systemic radial velocity of the source star measured from a spectrum taken while the source was highly magnified. Our measurement of $v_{\rm LSR} = 53 \pm 1$ km s$^{-1}$ does not definitively resolve the nature of the source, but supports the general conclusion that the source is in the bulge. 
    Although in this case the measured radial velocity was inconclusive, this work demonstrates that even a low signal-to-noise spectrum can provide useful information for characterizing microlensing source stars.
\end{abstract}

\maketitle

\section{Introduction}

One of the most fundamental assumptions in microlensing is that the
source star is in the bulge at $\sim 8$ kpc. This was the assumption in the
first calculations of the microlensing optical depth
\citep{Griest91,Paczynski91}. It also shows up in calculations of the
physical properties of the lens star, in which it is assumed that the
source is at the mean distance of the clump and experiences the same
amount of reddening \citep[e.g.][]{An02}. However, just as it rapidly became accepted that lens stars could come from the
bulge population as well as the disk
\citep{Udalski94,Alcock97,KiragaPaczynski94}, by \citet{Popowski01}
it was general knowledge that disk stars could also contribute to the population of
sources\footnote{In fact, \citet{Popowski01}
  attribute the idea of disk-disk lensing to
  \citet{KiragaPaczynski94}, but there is no evidence that
  \citet{KiragaPaczynski94} considered disk stars as
  sources.}. 
Disk
stars are likely to make up a few percent of microlensing sources,  but the true fraction of
events arising from disk sources remains unmeasured.
  
One way to distinguish disk sources from bulge sources is through
multi-band observations of microlensing events, which allow a
measurement of the source magnitude and color. In many cases, the
source's location on a color-magnitude diagram (CMD) can show whether
it comes from the disk or the bulge populations. However, sometimes
its CMD location is ambiguous, in which case more information is
needed. 

The work of \citet{Bensby13} suggests that measuring the radial
velocity of the source star can also provide information about which
population the source belongs to. They show that radial velocity measurements for their ensemble of 58 microlensing sources is broadly consistent
with a population of stars in the bulge. It should also be possible to use radial velocities to make inferences about individual sources. Given the velocity dispersions of
the disk ($\sigma = 39.7\pm 0.7$ km s$^{-1}$ \citealt{Sharma14}) and
the bulge ($\sigma = 93.5\pm 3.9$ km s$^{-1}$ \citealt{Ness13}), a
measurement of $\sim 100$ km s$^{-1}$ is strong evidence for a bulge
source. A low velocity (e.g. $\lesssim 30$ km s$^{-1}$) is more ambiguous because of the abundance of bulge stars, but might be combined with other evidence to demonstrate a disk source. Furthermore, in contrast to the \citet{Bensby13} work, which
also measured the detailed chemical abundances of stars, a
signal-to-noise (S/N) of only a few will yield an RV with sufficient precision to determine if it is in the bulge.

In practice, if the source location in the CMD is ambiguous, the
source itself will be a faint dwarf or subgiant star ($I \gtrsim
18.5$, for an apparent clump magnitude of $I_{\rm cl} \sim 16$). Thus,
a radial velocity measurement is possible with a large (8m-class)
telescope if the source dominates the light of the combined
source+blend. This can happen if the blend is very faint or if the
source is highly magnified. For \citet{Bensby13}, the goal for a
spectrum was S/N of $\sim 50$ with two hours of exposure time for a
dwarf source. This led to a requirement that the magnified sources
were $I\lesssim 16$, and therefore magnified by a factor of $A \gtrsim
50$. In contrast, a rough estimate of the radial velocity has much less stringent requirements on S/N and can therefore tolerate a fainter magnitude and a much lower magnification, thereby significantly increasing the number of sources for which such a measurement is possible.

In this paper, we present the analysis of a low S/N spectrum of
\thisevent\ to measure its radial velocity. \thisevent\ was originally
published by \citet{Street16} who showed that the lens consisted of a
star with a $q=1.7\times 10^{-4}$ planet. The interpretation
of this event suffers from multiple degeneracies. First, it
suffers from the well-known microlensing degeneracies for close and
wide binary solutions ($s\rightarrow s^{-1}$) and the four-fold
parallax degeneracy \citet{Gould94}. This leads to eight possible
solutions for the light curve. In all cases, this leads to a similar
value of $q$ and the difference in the magnitude of the parallax
($\pi_{\rm E}$) is not large \citep[see ][]{GouldYee12}. However, the
correct physical interpretation is complicated by the ambiguity in the
source location.

From their analysis of the CMD, \citet{Street16} find the position of
the source is consistent with both a subgiant in the bulge or a dwarf
star in the disk. The difference between these two options changes the determination of the source distance $D_{\rm S}$, which is used to calculate the distance to the system, $D_{\rm L}$:
\begin{equation}
     D_{\rm L}^{-1}  = \frac{\pi_{\rm E}\theta_{\rm E}}{\rm AU} + D_{\rm S}^{-1}.
    \label{eqn:dist}
\end{equation}
where $\theta_{\rm E}$ is the Einstein radius. The lens distance in turn affects the calculation of the physical star-planet projected separation, $r_{\perp}$ from
\begin{equation}
    s \equiv \frac{r_{\perp}}{\pi_{\rm E}\theta_{\rm E}}
\end{equation}
where $s$ is an observable of the microlensing light curve.
This problem was recognized while the event was still magnified. Thus, we obtained a spectrum of \thisevent S to measure the radial
velocity of the source star to attempt to resolve this ambiguity.

{\section{Data and Observations}\label{sec:data}}

\thisevent\ was observed with the Magellan Inamori Kyocera Echelle (MIKE) spectrometer \citep{Bernstein03} on the Magellan Clay telescope on the night of 2015 July 4. At the time of observations (HJD$^{\prime}=$HJD$-2450000.=7208.55$), the source was magnified by a factor of 16.7 and had an I magnitude of 16.6 \citep{Street16}. As noted in \citet{Street16}, the event had negative blending, so the source light dominates the spectrum.

The spectrum was obtained with a 1800s exposure taken in 0.7" seeing using the 0.7" slit and 2x2 binning. ThAr exposures were taken before and after the observation and spectra of bright B stars were used as flats. The spectrum of the source was reduced using the standard MIKE pipeline \citep{Kelson00,Kelson03}\footnote{http://code.obs.carnegiescience.edu/mike}. The pipeline did not produce the blue channel of the spectrum due to low signal.

{\section{Radial Velocity}\label{sec:rv}}

We determine the radial velocity (RV) of the source by performing a least squares fit of model spectra to the source star spectrum. 
\citet{Street16} find that the source star is consistent with both a subgiant star in the bulge and a main-sequence dwarf in the disk. 
To account for either scenario, we extract RVs using \texttt{PHOENIX} model spectra \citep{Husser2013}\footnote{http://phoenix.astro.physik.uni-goettingen.de/} for both scenarios and compare the results. 
Table \ref{tbl:stellar_props} gives the pertinent stellar properties for the models. 
The ``Reported'' values of absolute $I$-band magnitude, $M_I$, were determined using photometery, and the values of the surface gravity, log $g_*$, were determined from solar metallicity isochrones by \citet{Street16}. 
The ``Model Values'' are the adopted values for the \texttt{PHOENIX} model spectra based on their grid of available spectra. 
We derived the stellar effective temperature from $M_I$, and the log $g_*$ was simply rounded to the nearest value. 
We assume no $\alpha$-element enhancement in the source star. 
We convert the wavelengths of the models to in-air wavelengths through the prescription from \citet{Husser2013}. 

\begin{figure}
    \centering
    \includegraphics[width=\textwidth]{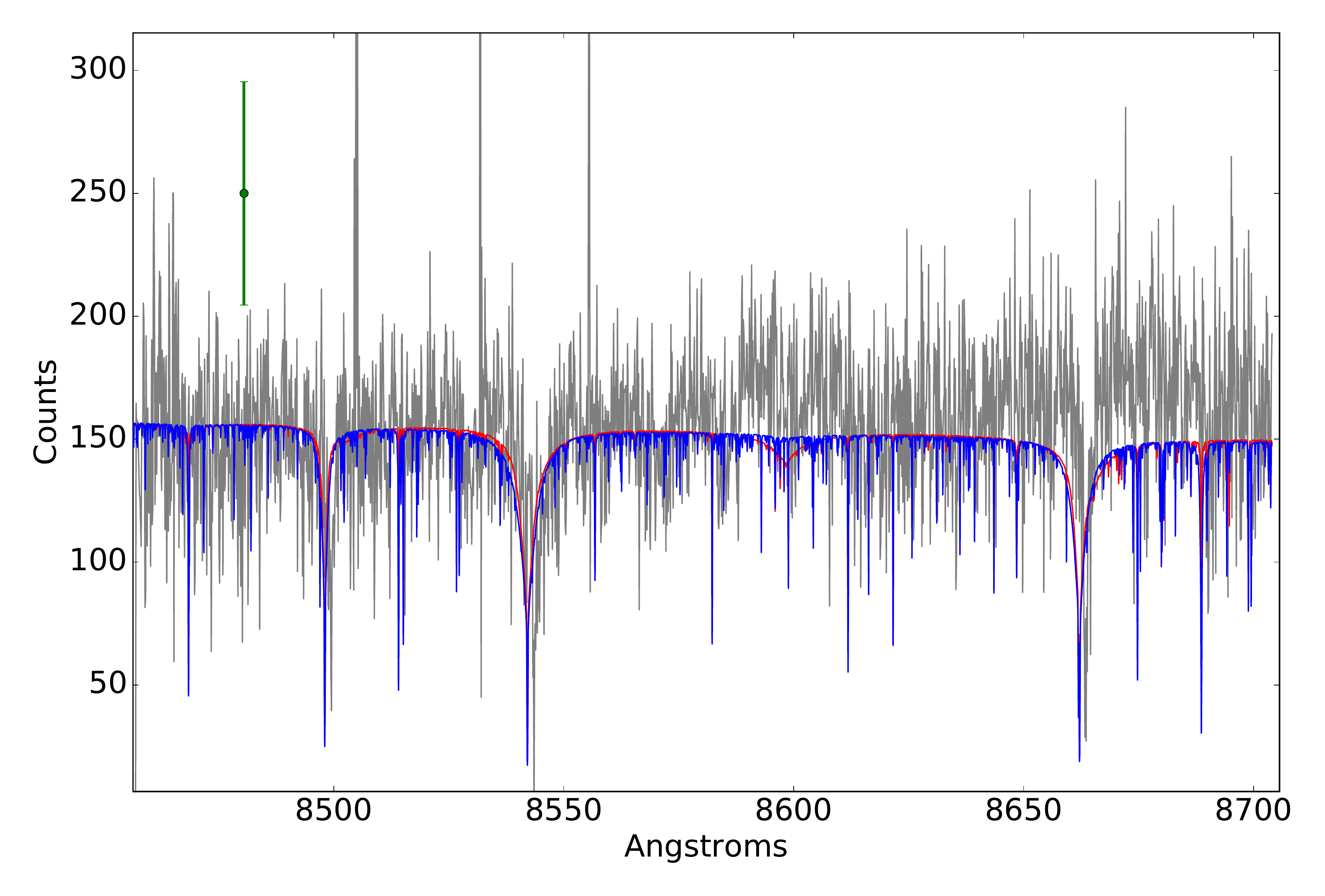}
    \caption{Spectral order 3 of \thisevent S containing the Ca I triplet (grey). Overplotted are the two \texttt{PHOENIX} models \citep{Husser2013}: main-sequence, disk source (blue) and subgiant, bulge source (red). The models are plotted at wavelengths in air, with no Doppler shifts or rebinning applied. A false point with the median error is plotted in green in the upper left.}
    \label{fig:spec_plot}
\end{figure}

The source star is faint, and the data are noisy, as shown in Figure \ref{fig:spec_plot}. 
We choose to only extract the Doppler shift from orders of the spectrum with a median SNR $>$2, which left us with 8 of the 34 available orders from the red channel of the MIKE data. 
For each order, the first and last 104 pixels were discarded, and the remaining pixels were divided into 364 pixel chunks ($\sim 45 $\AA) for which we could extract the Doppler shifts individually. 
The orders are masked through iterative 3$\sigma$-clipping for values greater than the median counts in the order. All bins with a value less than zero were also masked.

The general procedure for fitting the models to the data starts by convolving the model with the instrumental response (IR) of the instrument using \texttt{FFTCONVOLVE} from \texttt{SciPy} \citep{Jones2001}. 
We model the IR as a simple Gaussian with a width $w$. 
We then apply a Doppler shift to the wavelengths of the model by multiplying them by a factor (1+$z$). 
Next, we multiply the model by a quadratic factor in wavelength, $\lambda$, of the form  $f = c_0 + c_1\lambda + c_2\lambda^2$, where the $c_i$'s are free parameters. 
This allows broad features of the spectral chunk's continuum to be fit. 
Finally, we rebin the model spectra onto the same grid as the data using the \texttt{PySynphot} library \citep{lim2015}. 
The chi-squared value
\begin{equation}
    \chi^2 = \sum_i \frac{(m(\lambda_i,z,w,f) - d_i)^2}{\sigma_i^2}
\end{equation}
is minimized using the Python package \texttt{LMFIT} \citep{newville2014}. 
Here, $m(\lambda_i)$ is the model flux at the the $i$th wavelength, $d_i$ is the observed flux at the $i$th wavelength, and $\sigma_i$ is the error reported by the MIKE reduction pipeline for $d_i$.

Some orders lack the spectral features necessary to produce a good fit of the IR of the spectrograph, thus we broke the fitting process into two steps. 
In the first iteration, we fit all of the chunks simultaneously with the same Doppler shift and IR width. 
This allows the IR to be informed by all the orders but have strong absorption lines such as the Ca II triplet make up for the lack of features elsewhere. 
The difference between the widths of the IR measured by comparison to the two different model spectra was insignificant. 
In the second iteration, we fit the shift of the chunks independently but with the IR width fixed to that found in the previous step. 
This way, each of the spectral chunks had their Doppler shifts fit pseudo-independently, assuming the IR does not vary strongly with wavelength. 

The RVs and their uncertainties are given in Table \ref{tbl:rv_table1}.
We use a mixture model to examine the likelihood of the data and the mean value of the RV \citep{sivia1996}. 
This mixture model accounts for all data present by assuming points were drawn from either one of two distributions that share a mean: one with a narrow spread and the other wide. 
This way, we can reach a value that is informed by as much of the data as possible without assumptions on their quality. 
The individual probability for a data point is 
\begin{equation}
    \text{Pr}(D_n|\mu_{RV},\sigma_\mu,\beta,\gamma)=\frac{1}{\sigma_\mu\sqrt{2\pi}}\left((1-\beta)\exp{\frac{-(\mu_{RV}-D_n)^2}{2\sigma_\mu^2}}+\frac{\beta}{\gamma}\exp{\frac{-(\mu_{RV}-D_n)^2}{2\sigma_\mu^2\gamma^2}}\right).
\end{equation}
given the RV datum $D_n$ is drawn from one of the Gaussian distributions. 
The mean radial velocity, $\mu_{RV}$, is the center for both Gaussian distributions. 
The narrow distribution has a width of $\sigma_{\mu}$, while the wide distribution has a the same width scaled by a factor $\gamma$. 
The distributions must be complimentary in the sense that each RV measurement must be drawn from one distribution or the other; $\beta$ is the fraction of measurements from the wide distribution. 
The likelihood function then becomes the total product of all these probabilities,
\begin{equation}
\mathcal{L}(\mu_{RV},\sigma_\mu,\beta,\gamma|\mathbf{D_N}) = \prod_{n=0}^N \text{Pr}(D_n|\mu_{RV},\sigma_\mu,\beta,\gamma).
\end{equation}
We maximize the logarithm of this likelihood (using \texttt{LMFIT}), as it is the same as maximizing the likelihood function itself.  

\begin{figure}
    \centering
    \includegraphics[scale=0.4]{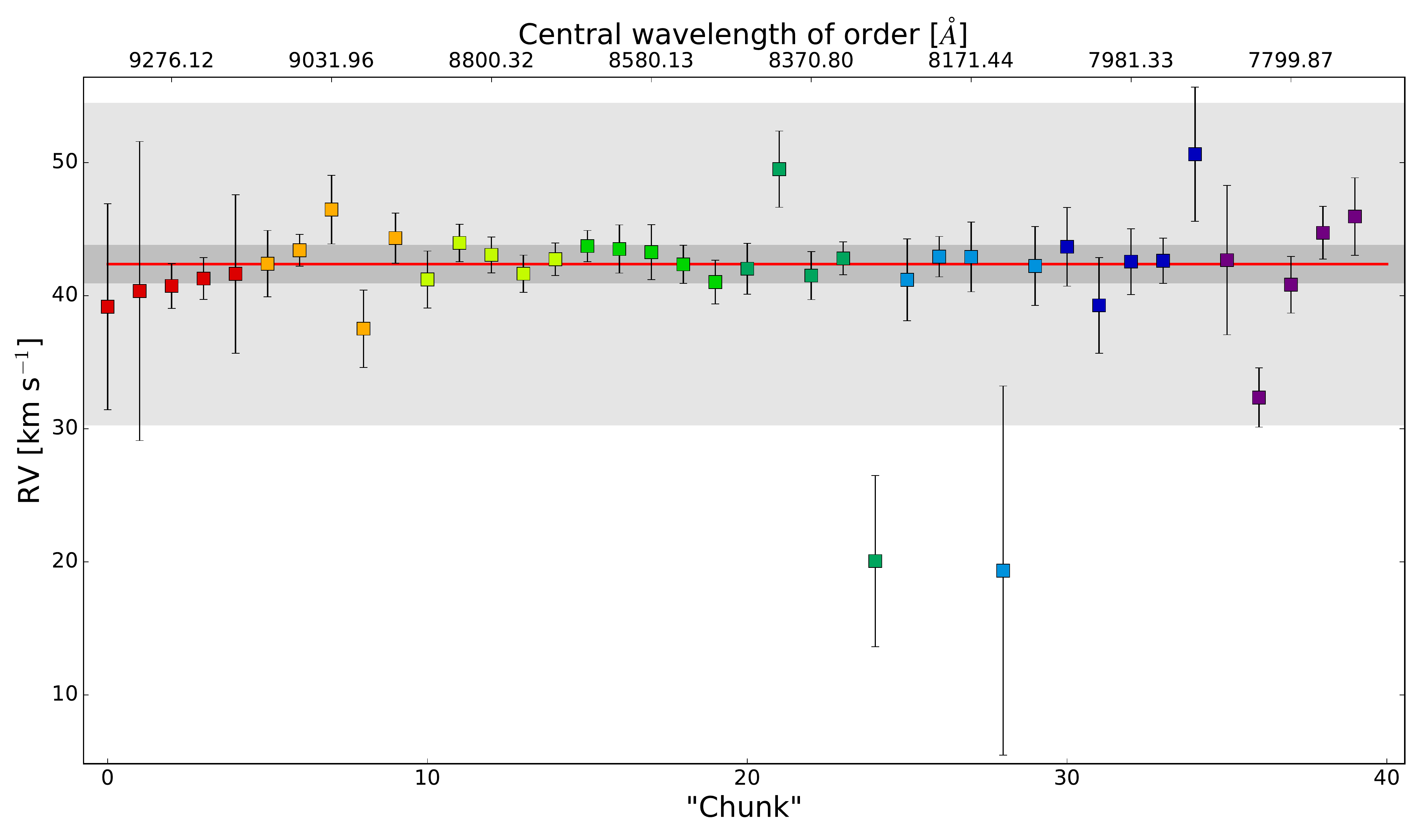}
    \caption{The chunk number versus the measured RVs from the subgiant model spectrum. Each order, starting from the first order in the spectrum, is divided into five chunks; points from the same order have the same color. The red line is the determined mean from the maximum likelihood, $\mu_{RV}$, at 42.4 km s$^{-1}$. The dark, inner grey band shows $\sigma_{\mu}$ from the maximum likelihood analysis, while the light, outer grey box indicates $\sigma_{\mu}*\gamma$. }
    \label{fig:rv_plot_sg}
\end{figure}

We use the IDL module {\texttt{BARYCORR}} from \citet{Wright14} to determine a barycentric correction of -7.2 km s$^{-1}$. 
The module is orders of magnitude more precise than our measurement. 
Therefore, we consider any error in the barycentric correction to be negligible compared with the uncertainty in our measured RV. 
Table \ref{tbl:maxlike} contains the results from both models after combining the RVs with the barycentric correction and Figure \ref{fig:rv_plot_sg} shows the measured RV as a function of order for the subgiant reference model spectrum. 

We find a heliocentric RV of 42.8 $\pm$ 1.4 km s$^{-1}$ using the main-sequence dwarf model spectrum and 42.4 $\pm$ 1.5 km s$^{-1}$ using the subgiant model spectrum after combining the results from the maximum likelihood analysis with the barycentric correction. 
The $\beta$ parameter for both scenarios indicated less than one third of the RV points were from the broader distribution.
The agreement between the RV measurements show that they are insensitive to variations in the reference model spectrum.
We adopt $42.4\pm1.5$ km s$^{-1}$ (the subgiant model value) as our measurement of the heliocentric radial velocity for the source star.

{\section{Discussion}\label{sec:conclude}}

To test whether or not the source is in the disk or the bulge, we must account for the motion of the Sun relative to the local standard of rest, $10.96\pm0.14$ km s$^{-1}$ \citep{Sharma14}\footnote{\citet{Sharma14} give asymmetric uncertainties for $U_{\odot}$, but for simplicity, we adopt the larger of the two values.}. When including the solar motion, we  find that the source is moving towards the galactic center at
\begin{equation}
    v_{\rm RV, LSR} = 53.4 \pm 1.5\, \mathrm{km\, s}^{-1}.
\end{equation}

Unfortunately, this velocity does not clearly distinguish between the disk and bulge scenarios for the source star. 
The measured velocity of the source star is consistent with either population. 
Because bulge stars are at least a factor of $\sim 4$ more abundant due to both volume and stellar density effects \citep{HanGould03}, we conclude that the source star is more likely to be in the bulge. 
If this is the case, then the distance to the lens system is $D_{\rm L} = 3.3\,$ kpc \citep{Street16}. 
The only remaining avenue to resolve this degeneracy would be to measure the proper motion of the source, which requires high-resolution imaging in $\sim 10\,$ years \citep{Street16}.

Although we were unlucky in the case of \thisevent S, we have shown that a spectrum with signal-to-noise of $\sim3$ is sufficient to measure the radial velocity of a source star. 
A somewhat higher S/N ratio would allow for a measurement of the spectral type of the source, but was not possible in this case. 
Such measurements require real-time triggering of target-of-opportunity spectra, but the requirements are much less stringent than for previous applications \citep{Bensby13, Bensby16}, leading to a larger window for observations. 
If such a spectrum can be obtained, the radial velocity of the source star and its spectral type provide additional information for determining the nature of the source. 
OGLE-2013-BLG-0911 is one event for which the ensemble of evidence, including the measured radial velocity, indicates that the source is indeed a member of the disk rather than the bulge \citep{Bensby16}. 
A systematic analysis of such sources could lead to a direct measurement of the relative contributions of bulge and disk sources to the observed microlensing event rates and optical depth.


\acknowledgements{We thank Ian Czekala for obtaining the spectrum of this event with MIKE, Christian Johnson for a preliminary examination of the spectrum, John A. Johnson and Jason Eastman for guidance on the RV extraction process, and Andrew Gould for comments on the manuscript. Work by SAJ was funded through NASA EPSCOR grant NNX13AM97A and NSF grant 1516242. Work by J.C.Y. was performed in part under contract with
the California Institute of Technology (Caltech)/Jet Propulsion
Laboratory (JPL) funded by NASA through the Sagan Fellowship Program
executed by the NASA Exoplanet Science Institute.}

\bibliography{references}

\begin{thebibliography}{}
\expandafter\ifx\csname natexlab\endcsname\relax\def\natexlab#1{#1}\fi

\bibitem[{{Alcock} {et~al.}(1997){Alcock}, {Allsman}, {Alves}, {Axelrod},
  {Bennett}, {Cook}, {Freeman}, {Griest}, {Guern}, {Lehner}, {Marshall},
  {Park}, {Perlmutter}, {Peterson}, {Pratt}, {Quinn}, {Rodgers}, {Stubbs}, \&
  {Sutherland}}]{Alcock97}
{Alcock}, C., {Allsman}, R.~A., {Alves}, D., {et~al.} 1997, \apj, 479, 119

\bibitem[{{An} {et~al.}(2002){An}, {Albrow}, {Beaulieu}, {Caldwell}, {DePoy},
  {Dominik}, {Gaudi}, {Gould}, {Greenhill}, {Hill}, {Kane}, {Martin},
  {Menzies}, {Pogge}, {Pollard}, {Sackett}, {Sahu}, {Vermaak}, {Watson}, \&
  {Williams}}]{An02}
{An}, J.~H., {Albrow}, M.~D., {Beaulieu}, J.-P., {et~al.} 2002, \apj, 572, 521

\bibitem[{{Bensby} {et~al.}(2013){Bensby}, {Yee}, {Feltzing}, {Johnson},
  {Gould}, {Cohen}, {Asplund}, {Mel{\'e}ndez}, {Lucatello}, {Han}, {Thompson},
  {Gal-Yam}, {Udalski}, {Bennett}, {Bond}, {Kohei}, {Sumi}, {Suzuki}, {Suzuki},
  {Takino}, {Tristram}, {Yamai}, \& {Yonehara}}]{Bensby13}
{Bensby}, T., {Yee}, J.~C., {Feltzing}, S., {et~al.} 2013, \aap, 549, A147

\bibitem[{{Bensby}(2016)}]{Bensby16}
{Bensby}, T.~e. 2016, in prep

\bibitem[{{Bernstein} {et~al.}(2003){Bernstein}, {Shectman}, {Gunnels},
  {Mochnacki}, \& {Athey}}]{Bernstein03}
{Bernstein}, R., {Shectman}, S.~A., {Gunnels}, S.~M., {Mochnacki}, S., \&
  {Athey}, A.~E. 2003, in \procspie, Vol. 4841, Instrument Design and
  Performance for Optical/Infrared Ground-based Telescopes, ed. M.~{Iye} \&
  A.~F.~M. {Moorwood}, 1694--1704

\bibitem[{{Gould}(1994)}]{Gould94}
{Gould}, A. 1994, \apjl, 421, L75

\bibitem[{{Gould} \& {Yee}(2012)}]{GouldYee12}
{Gould}, A., \& {Yee}, J.~C. 2012, \apjl, 755, L17

\bibitem[{{Griest} {et~al.}(1991){Griest}, {Alcock}, {Axelrod}, {Bennett},
  {Cook}, {Freeman}, {Park}, {Perlmutter}, {Peterson}, {Quinn}, {Rodgers},
  {Stubbs}, \& {MACHO Collaboration}}]{Griest91}
{Griest}, K., {Alcock}, C., {Axelrod}, T.~S., {et~al.} 1991, \apjl, 372, L79

\bibitem[{{Han} \& {Gould}(2003)}]{HanGould03}
{Han}, C., \& {Gould}, A. 2003, \apj, 592, 172

\bibitem[{{Husser} {et~al.}(2013){Husser}, {Wende-von Berg}, {Dreizler},
  {Homeier}, {Reiners}, {Barman}, \& {Hauschildt}}]{Husser2013}
{Husser}, T.-O., {Wende-von Berg}, S., {Dreizler}, S., {et~al.} 2013, \aap,
  553, A6

\bibitem[{Jones {et~al.}(2001)Jones, Oliphant, Peterson, {et~al.}}]{Jones2001}
Jones, E., Oliphant, T., Peterson, P., {et~al.} 2001, {SciPy}: Open source
  scientific tools for {Python}

\bibitem[{{Kelson}(2003)}]{Kelson03}
{Kelson}, D.~D. 2003, \pasp, 115, 688

\bibitem[{{Kelson} {et~al.}(2000){Kelson}, {Illingworth}, {van Dokkum}, \&
  {Franx}}]{Kelson00}
{Kelson}, D.~D., {Illingworth}, G.~D., {van Dokkum}, P.~G., \& {Franx}, M.
  2000, \apj, 531, 159

\bibitem[{{Kiraga} \& {Paczynski}(1994)}]{KiragaPaczynski94}
{Kiraga}, M., \& {Paczynski}, B. 1994, \apjl, 430, L101

\bibitem[{Lim {et~al.}(2015)Lim, Diaz, \& Laidler}]{lim2015}
Lim, P., Diaz, R., \& Laidler, V. 2015, PySynphot User’s Guide

\bibitem[{{Ness} {et~al.}(2013){Ness}, {Freeman}, {Athanassoula},
  {Wylie-de-Boer}, {Bland-Hawthorn}, {Asplund}, {Lewis}, {Yong}, {Lane},
  {Kiss}, \& {Ibata}}]{Ness13}
{Ness}, M., {Freeman}, K., {Athanassoula}, E., {et~al.} 2013, \mnras, 432, 2092

\bibitem[{Newville {et~al.}(2014)Newville, Stensitzki, Allen, \&
  Ingargiola}]{newville2014}
Newville, M., Stensitzki, T., Allen, D.~B., \& Ingargiola, A. 2014, {LMFIT:
  Non-Linear Least-Square Minimization and Curve-Fitting for Python},
  doi:10.5281/zenodo.11813

\bibitem[{{Paczynski}(1991)}]{Paczynski91}
{Paczynski}, B. 1991, \apjl, 371, L63

\bibitem[{{Popowski} {et~al.}(2001){Popowski}, {Alcock}, {Allsman}, {Alves},
  {Axelrod}, {Becker}, {Bennett}, {Cook}, {Drake}, {Freeman}, {Geha}, {Griest},
  {Lehner}, {Marshall}, {Minniti}, {Nelson}, {Peterson}, {Pratt}, {Quinn},
  {Stubbs}, {Sutherland}, {Tomaney}, {Vandehei}, \& {Welch}}]{Popowski01}
{Popowski}, P., {Alcock}, C., {Allsman}, R.~A., {et~al.} 2001, in Astronomical
  Society of the Pacific Conference Series, Vol. 239, Microlensing 2000: A New
  Era of Microlensing Astrophysics, ed. J.~W. {Menzies} \& P.~D. {Sackett}, 244

\bibitem[{{Sharma} {et~al.}(2014){Sharma}, {Bland-Hawthorn}, {Binney},
  {Freeman}, {Steinmetz}, {Boeche}, {Bienaym{\'e}}, {Gibson}, {Gilmore},
  {Grebel}, {Helmi}, {Kordopatis}, {Munari}, {Navarro}, {Parker}, {Reid},
  {Seabroke}, {Siebert}, {Watson}, {Williams}, {Wyse}, \& {Zwitter}}]{Sharma14}
{Sharma}, S., {Bland-Hawthorn}, J., {Binney}, J., {et~al.} 2014, \apj, 793, 51

\bibitem[{Sivia(1996)}]{sivia1996}
Sivia, D. 1996, Maximum Entropy and Bayesian Methods, 131

\bibitem[{{Street} {et~al.}(2016){Street}, {Udalski}, {Calchi Novati},
  {Hundertmark}, {Zhu}, {Gould}, {Yee}, {Tsapras}, {Bennett}, {RoboNet
  Project}, {Consortium}, {J{\o}rgensen}, {Dominik}, {Andersen}, {Bachelet},
  {Bozza}, {Bramich}, {Burgdorf}, {Cassan}, {Ciceri}, {D'Ago}, {Dong}, {Evans},
  {Gu}, {Harkonnen}, {Hinse}, {Horne}, {Figuera Jaimes}, {Kains}, {Kerins},
  {Korhonen}, {Kuffmeier}, {Mancini}, {Menzies}, {Mao}, {Peixinho}, {Popovas},
  {Rabus}, {Rahvar}, {Ranc}, {Tronsgaard Rasmussen}, {Scarpetta}, {Schmidt},
  {Skottfelt}, {Snodgrass}, {Southworth}, {Steele}, {Surdej}, {Unda-Sanzana},
  {Verma}, {von Essen}, {Wambsganss}, {Wang}, {Wertz}, {OGLE Project},
  {Poleski}, {Pawlak}, {Szyma{\'n}ski}, {Skowron}, {Mr{\'o}z}, {Koz{\l}owski},
  {Wyrzykowski}, {Pietrukowicz}, {Pietrzy{\'n}ski}, {Soszy{\'n}ski}, {Ulaczyk},
  {Spitzer Team}, {Beichman}, {Bryden}, {Carey}, {Gaudi}, {Henderson}, {Pogge},
  {Shvartzvald}, {The MOA Collaboration}, {Abe}, {Asakura}, {Bhattacharya},
  {Bond}, {Donachie}, {Freeman}, {Fukui}, {Hirao}, {Inayama}, {Itow},
  {Koshimoto}, {Li}, {Ling}, {Masuda}, {Matsubara}, {Muraki}, {Nagakane},
  {Nishioka}, {Ohnishi}, {Oyokawa}, {Rattenbury}, {Saito}, {Sharan},
  {Sullivan}, {Sumi}, {Suzuki}, {Tristram}, {Wakiyama}, {Yonehara}, {KMTNet
  Modeling Team}, {Han}, {Choi}, {Park}, {Jung}, \& {Shin}}]{Street16}
{Street}, R.~A., {Udalski}, A., {Calchi Novati}, S., {et~al.} 2016, \apj, 819,
  93

\bibitem[{{Udalski} {et~al.}(1994){Udalski}, {Szymanski}, {Stanek}, {Kaluzny},
  {Kubiak}, {Mateo}, {Krzeminski}, {Paczynski}, \& {Venkat}}]{Udalski94}
{Udalski}, A., {Szymanski}, M., {Stanek}, K.~Z., {et~al.} 1994, \actaa, 44, 165

\bibitem[{{Wright} \& {Eastman}(2014)}]{Wright14}
{Wright}, J.~T., \& {Eastman}, J.~D. 2014, \pasp, 126, 838

\end{thebibliography}

\pagebreak

\begin{deluxetable}{c|cc|ccc}
\centering

\tablecaption{Properties for \texttt{PHOENIX} Model Spectra\label{tbl:stellar_props}}
\tablehead{\colhead{}
&\multicolumn{2}{|c|}{Reported}&\multicolumn{3}{c}{Model Values}\\
\colhead{Model}&\multicolumn{1}{|c}{$M_I$}&\multicolumn{1}{c|}{log $g_*$}&\colhead{$T_{\rm eff}$ [K]}&\colhead{log $g_*$}&\colhead{$\alpha$ enhancement}}
\startdata
Main sequence&4.15&4.4&5500&4.5&0.0\\
Subgiant&2.96&3.9&6300&4.0&0.0\\
\enddata
\end{deluxetable}
\begin{deluxetable}{cccc|cc|cc}
\tablecaption{Measured RVs\label{tbl:rv_table1}}
\tablehead{
\multicolumn{4}{c}{}&\multicolumn{2}{|c|}{MS}&\multicolumn{2}{c}{Subgiant}\\
\colhead{Order}&\colhead{Chunk}&\colhead{$\lambda_{\rm min}$}&\colhead{$\lambda_{\rm max}$}&\multicolumn{1}{|c}{$RV$}&\multicolumn{1}{c|}{$\sigma_{RV}$}&\multicolumn{1}{|c}{$RV$}&\multicolumn{1}{c}{$\sigma_{RV}$}}
\startdata
0 & 4 & 9348.38 & 9396.38 & 34.5 & 6.2  & 39.2 & 7.7\\
0 & 3 & 9300.25 & 9348.25 & 20.2 & 8.6  & 40.4 & 11.2\\
0 & 2 & 9252.12 & 9300.12 & 45.3 & 1.9  & 40.7 & 1.7\\
0 & 1 & 9203.99 & 9251.99 & 42.3 & 2.3  & 41.3 & 1.6\\
0 & 0 & 9155.86 & 9203.86 & 43.2 & 4.0  & 41.6 & 6.0\\
\hline
1 & 4 & 9102.32 & 9149.06 & 43.0 & 3.4  & 42.4 & 2.5\\
1 & 3 & 9055.45 & 9102.19 & 44.4 & 1.9  & 43.4 & 1.2\\
1 & 2 & 9008.59 & 9055.33 & 43.5 & 2.5  & 46.5 & 2.6\\
1 & 1 & 8961.72 & 9008.46 & 38.4 & 3.1  & 37.5 & 2.9\\
1 & 0 & 8914.86 & 8961.59 & 44.3 & 2.4  & 44.3 & 1.9\\
\hline
2 & 4 & 8868.88 & 8914.43 & 41.2 & 3.4  & 41.2 & 2.1\\
2 & 3 & 8823.22 & 8868.76 & 43.8 & 1.5  & 44.0 & 1.4\\
2 & 2 & 8777.55 & 8823.09 & 43.1 & 1.6  & 43.1 & 1.4\\
2 & 1 & 8731.88 & 8777.42 & 42.2 & 1.5  & 41.7 & 1.4\\
2 & 0 & 8686.21 & 8731.76 & 44.2 & 1.4  & 42.7 & 1.2\\
\hline
3 & 4 & 8646.99 & 8691.40 & 43.2 & 1.3  & 43.7 & 1.2\\
3 & 3 & 8602.46 & 8646.87 & 48.5 & 1.9  & 43.5 & 1.8\\
3 & 2 & 8557.93 & 8602.34 & 43.4 & 2.2  & 43.3 & 2.1\\
3 & 1 & 8513.40 & 8557.81 & 42.5 & 1.7  & 42.4 & 1.4\\
3 & 0 & 8468.87 & 8513.28 & 41.5 & 1.9  & 41.0 & 1.7\\
\hline
4 & 4 & 8436.04 & 8479.37 & 41.6 & 1.9  & 42.0 & 1.9\\
4 & 3 & 8392.59 & 8435.92 & 48.8 & 2.3  & 49.5 & 2.9\\
4 & 2 & 8349.14 & 8392.47 & 43.3 & 1.7  & 41.5 & 1.8\\
4 & 1 & 8305.69 & 8349.02 & 42.5 & 1.5  & 42.8 & 1.2\\
4 & 0 & 8262.24 & 8305.57 & 22.4 & 5.6  & 20.1 & 6.4\\
\hline
5 & 4 & 8235.12 & 8277.42 & 41.1 & 3.9  & 41.2 & 3.1\\
5 & 3 & 8192.70 & 8235.00 & 43.3 & 1.8  & 42.9 & 1.5\\
5 & 2 & 8150.29 & 8192.59 & 43.6 & 2.8  & 42.9 & 2.6\\
5 & 1 & 8107.87 & 8150.17 & 65.8 & 7.6  & 19.3 & 13.9\\
5 & 0 & 8065.46 & 8107.76 & 41.8 & 2.7  & 42.2 & 3.0\\
\hline
6 & 4 & 8043.53 & 8084.85 & 40.6 & 2.6  & 43.7 & 3.0\\
6 & 3 & 8002.10 & 8043.42 & 40.2 & 4.0  & 39.3 & 3.6\\
6 & 2 & 7960.67 & 8001.99 & 41.6 & 2.7  & 42.6 & 2.5\\
6 & 1 & 7919.24 & 7960.56 & 42.3 & 2.0  & 42.6 & 1.7\\
6 & 0 & 7877.81 & 7919.13 & 46.9 & 5.3  & 50.6 & 5.0\\
\hline
7 & 4 & 7860.66 & 7901.04 & 44.6 & 9.7  & 42.7 & 5.6\\
7 & 3 & 7820.17 & 7860.55 & 34.5 & 2.6  & 32.4 & 2.2\\
7 & 2 & 7779.68 & 7820.06 & 40.7 & 2.3  & 40.8 & 2.1\\
7 & 1 & 7739.19 & 7779.57 & 47.0 & 3.1  & 44.7 & 2.0\\
7 & 0 & 7698.70 & 7739.08 & 44.4 & 3.3  & 46.0 & 2.9\\
\enddata
\tablecomments{The values of the extracted radial velocities, $RV$, and the errors, $\sigma_{RV}$, from the individual fitting of the spectral chunks using the main-sequence (MS) and sub-giant (Sub) models. Also included are the minimum and maximum wavelengths, $\lambda_{min}$ and $\lambda_{max}$, for each spectral chunk.}
\end{deluxetable}

\begin{deluxetable}{c|ccccc}
\tablecaption{Final RVs\label{tbl:maxlike}}
\tablehead{
\colhead{Model}&\colhead{$\mu_{RV}$ [km s$^{-1}$]}&\colhead{$\sigma_{\mu}$ [km s$^{-1}$]}&\colhead{$\beta$}&\colhead{$\gamma$}&\colhead{\# of Chunks}
}
\startdata
Main-sequence & 42.8 & 1.4 & 0.29 & 8.7 & 40\\
Subgiant      & 42.4 & 1.5 & 0.22 & 8.3 & 40\\
\enddata
\tablecomments{The maximized parameters of the likelihood function given the data from the individual fitting of the spectral chunks. The mean RV, $\mu_{RV}$, the standard deviation of the narrow distribution, $\sigma_{\mu}$, the percentage of data drawn from the broader distribution, $\beta$, and the multiplicative factor for $\sigma_{\mu}$ in the broader distribution, $\gamma$, are reported, as well as the number of spectral chunks fit. The RV values agree with in the error bounds, suggesting that the result is overall independent of the model used in fitting. }
\end{deluxetable}

\end{document}